\documentstyle [11pt]{article}
\def\journal#1#2#3#4{\ {\sl #1}\ \underline{\bf #2}, {#3}\  ({#4})}

\def\ActaPolonica{\journal{Acta \ Physica \ Polonica\/}}

\def\AnnPhys{\journal{Ann.\ Phys.}}
\def\CommMathPhys{\journal{Comm. \ Math. \ Phys.}}

\def\ibid{\journal{\em ibid.}}
\def\IJMPA{\journal{Int.\ J.\ Mod.\ Phys \ {\bf A}}}

\def\NPB{\journal{Nucl.\ Phys.\ {\bf B}}}

\def\PLB{\journal{Phys.\ Lett.\ {\bf B}}}
\def\PhysRev{\journal{Phys.\ Rev.}}

\def\PRD{\journal{Phys.\ Rev.\ {\bf D}}}
\def\PRL{\journal{Phys.\ Rev.\ Lett.}}
\def\PhysRept{\journal{Phys.\ Repts.}}

\def\RevModPhys{\journal{Rev.\ Mod.\ Phys.}}

\def\SovJNuclPhys{\journal{Sov.\ J.\ Nucl.\ Phys.}}

\def\ZPhysC{\journal{Z.\ Phys.\ C}}

\begin{document}

\newcommand{\be}{\begin{equation}}
\newcommand{\ee}{\end{equation}}
\newcommand{\barray}{\begin{eqnarray}}
\newcommand{\earray}{\end{eqnarray}}
\newcommand{\gn}{\mbox{$\gamma_{_{5}}$}}
\newcommand{\adag}{a^{\dagger}_{p,s}}
\newcommand{\atilde}{\tilde{a}^{}_{p,s}}
\newcommand{\atildedag}{\tilde{a}^{\dagger}_{-p,s}}
\newcommand{\bdag}{b^{\dagger}_{-p,s}}
\newcommand{\btilde}{\tilde{b}^{}_{-p,s}}
\newcommand{\btildedag}{\tilde{b}^{\dagger}_{-p,s}}
\newcommand{\apsbeta}{a^{\beta}_{p,s}}
\newcommand{\apsbetadag}{a_{-p,s}^{\beta\dagger}}
\newcommand{\adagbdag}{a^{\dagger}_{p,s} b^{\dagger}_{-p,s}}
\newcommand{\aps}{a^{}_{p,s}}
\newcommand{\bps}{b^{}_{-p,s}}
\newcommand{\bpsbeta}{b^{\beta}_{-p,s}}
\newcommand{\bpsbetadag}{b_{-p,s}^{\beta\dagger}}
\newcommand{\Apsdag}{A^{\dagger}_{p,s}}
\newcommand{\Bpsdag}{B^{\dagger}_{-p,s}}
\newcommand{\Aps}{A^{}_{p,s}}
\newcommand{\Apsbeta}{A^{\beta}_{p,s}}
\newcommand{\Apsbetadag}{A^{\beta\dagger}_{p,s}}
\newcommand{\Bps}{B^{}_{-p,s}}
\newcommand{\Bpsbeta}{B^{\beta}_{-p,s}}
\newcommand{\Bpsbetadag}{B^{\beta\dagger}_{-p,s}}
\newcommand{\ApL}{A^{}_{p,L}}
\newcommand{\BpL}{B^{}_{-p,L}}
\newcommand{\ApR}{A^{}_{p,R}}
\newcommand{\BpR}{B^{}_{-p,R}}
\newcommand{\BdagL}{B^{\dagger}_{-p,L}}
\newcommand{\BdagR}{B^{\dagger}_{-p,R}}
\newcommand{\eps}{\epsilon}
\newcommand{\go}{\left( \gamma_o - \vec{\gamma} \cdot \hat{n} \right)}
\newcommand{\abab}{a^{\dagger}_{p,L}\,b^{\dagger}_{-p,L}\,a^{\dagger}_{p,R}
                   \,b^{\dagger}_{-p,R}}
\newcommand{\alphai}{\alpha_{i}}
\newcommand{\limit}{\lim_{\Lambda^2 \rightarrow \infty}}
\newcommand{\p}{\vec{p}, p_o}
\newcommand{\poprime}{p_o^{\prime}}
\newcommand{\prodps}{\prod_{p,s}}
\newcommand{\prodp}{\prod_{p}}
\newcommand{\psibar}{\bar{\psi}}
\newcommand{\psibarpsi}{ < \bar{\psi} \, \psi
            > }
\newcommand{\PsibarPsi}{ < \bar{\Psi} \, \Psi
            > }
\newcommand{\psibeta}{\psi^{}_{\beta}}
\newcommand{\psibarbeta}{\bar{\psi}_{\beta}}
\newcommand{\psidag}{\psi^{\dagger}}
\newcommand{\psidagbeta}{\psi^{\dagger}_{\beta}}
\newcommand{\psiL}{\psi_{_{L}}}
\newcommand{\psiR}{\psi_{_{R}}}
\newcommand{\Q}{Q_{_{5}}}
\newcommand{\Qa}{Q_{_{5}}^{a}}
\newcommand{\Qij}{Q_{5}{}^{i}_{j}}
\newcommand{\Qbeta}{Q_{5}^{\beta}}
\newcommand{\Qabeta}{Q_{5}^{a\,\beta}}
\newcommand{\qqbar}{q\bar{q}}
\newcommand{\sumps}{\sum_{p,s}}
\newcommand{\TH}{\Theta}
\newcommand{\thetap}{\theta_{p}}
\newcommand{\thetapp}{\theta_{p}^{\prime}}
\newcommand{\costhetap}{\cos{\thetap}}
\newcommand{\sinthetap}{\sin{\thetap}}
\newcommand{\costthetap}{\cos{2\thetap}}
\newcommand{\sintthetap}{\sin{2\thetap}}
\newcommand{\costthetapp}{\cos{2\thetapp}}
\newcommand{\sintthetapp}{\sin{2\thetapp}}
\newcommand{\cosxi}{\cos{\xi}}
\newcommand{\cosxip}{\cos{\xi^{\prime}}}
\newcommand{\sinxi}{\sin{\xi}}
\newcommand{\sinxip}{\sin{\xi^{\prime}}}
\newcommand{\thetaset}{\{ \thetap  \}}
\newcommand{\thetapi}{\thetap{}_{i}}
\newcommand{\Tomega}{\frac{\Tprime}{\omega}}
\newcommand{\pomega}{\frac{p}{\omega}}
\newcommand{\Tprime}{T^{'}}
\newcommand{\Tprimesq}{T^{'2}}
\newcommand{\vac}{| vac \rangle}
\newcommand{\vacbeta}{| vac \rangle_{_{\beta}}}
\newcommand{\vecp}{\vec{p}}
\newcommand{\vecpdotx}{\vec{p} \cdot \vec{x}}
\newcommand{\vecx}{\vec{x}}
\newcommand{\x}{\vec{x},t}
\newcommand{\xip}{\xi^{\prime}}
\newcommand{\xPrime}{\vec{x} - \hat{n} ( t - t'), t'}
\newcommand{\y}{\vec{y}, y_o}

\thispagestyle{empty}

\vspace*{-.15in}

\hspace*{\fill}\fbox{CCNY-HEP-94-6}

\begin{center}
{\Large {\bf Chiral Morphing
\fnsymbol{footnote}\footnote
{\samepage \sl
\noindent \parbox[t]{5.5in}{ \noindent
        Invited contribution to the Marshak Memorial volume, edited
	by E.C.G. Sudarshan, to be published by World Scientific
	Publishing Co., Singapore, 1994.
                           }

}
}
}\\
\baselineskip 5mm
\ \\
Ngee-Pong Chang (npccc@cunyvm.cuny.edu)\\
Department of Physics\\
City College \& The Graduate School of City University of New York\\
New York, N.Y. 10031\\
\  \\
March 22, 1994  \\

\end{center}

\vspace*{-.15in}

\noindent\hspace*{\fill}\parbox[t]{4.5in}{
        \hspace*{\fill}{\bf Abstract}\hspace*{\fill} \\
        {\em
	Chiral symmetry undergoes a metamorphosis at $T_c$.
	For $T < T_c$, the usual Noether charge, $\Qa$, is dynamically
	broken by the vacuum.   Above $T_c$, chiral symmetry undergoes
	a subtle change, and the Noether charge \underline{{\em  morphs}}
	into $\Qbeta$, with the thermal vacuum now becoming invariant
	under $\Qbeta$.  \\

	This vacuum is however not invariant under the old $\Qa$
	transformations.
	As a result, the pion remains strictly massless at high $T$.
	The pion propagates in the early universe with a halo. \\

	New order parameters are proposed to probe the structure of
	the new thermal vacuum.

        }
                              }\hspace*{\fill} \\


\section{Prologue}
{\em
	It is a pleasure for me to dedicate this to the memory of
	Marshak, whose life work has revolved around the pion and
	chirality.  When he first came to City College as the
	eighth President in 1971, he not only plunged into his
	duties as president with characteristic energy and gusto,
	scheduling breakfast meetings with faculty and staff,
	but he also squeezed into his day an early ( 8 a.m. )
	theoretical physics seminar in the President's Conference Room
	during his first year at City College.

	In those days, it was with a sense of awe that we
	got to step into the halycon halls of administration for
	a round of physics seminars.  And promptly at the hour
	of $8$, Marshak would enter the Conference Room with the
	big oval table and the seminar would begin.  Some of
	the invited speakers had come from as far as sixty miles away!

	After Sakita returned from France, the seminar was moved back
	to the more normal hour of $2 \, p.m.$ on Friday.  By then, Marshak
	had gotten the hang of the presidency, and religiously found time
	to come to the seminar for the hour and half of physics with peace
	and tranquillity.

	His interest in quality and excellence did not flag through those
	days of NY City fiscal crisis, budget cuts and student unrest.
	In spite of it, he was able to set up the Sophie Davis biomedical
	program that has become one of the jewels of City College, the
	Levich institute of hydrodynamics, and many other initiatives that
	have had a lasting impact on City College.

	For the setting up of the Levich institute, Marshak acted swiftly
	upon hearing of the impending emigration of Benjamin Levich from
	Russia and pulled all strings both domestically and internationally
	to get Benjamin Levich to come to City College.  There were sensitive
	cold war constraints and other complications, and the deal involved
	Israel.
	And so on one Friday afternoon, Marshak entrusted Bunji and me with a
	hurried mission to intercept Yuval Ne'eman on one of his unannounced
	trips to New York.

	So there we were, standing at the exit hall of the Kennedy airport
	waiting for El Al passengers to clear customs.  We waited and waited.
	We checked with El Al to verify that Ne'eman was on the passenger
	list.  The El Al security quietly alerted Ne'eman to the fact that
	`two unidentified oriental strangers were standing in the hallway'.
	And so Ne'eman was asked to come out by
	a side entrance, and we were finally accosted by Ne'eman only after he
	recognized who we were.

	In those days, there was nothing that Marshak could not do
	as a President.  We at City College owe a lot to him for his
	having assumed the mantle of office after open admissions plunged
	the college into a crisis of identity.

	Amid all the turmoil, it was fitting that City College was the
	venue in 1977 for a 60th birthday celebration of Marshak as a
	truly distinguished physicist of international renown and stature,
	a scholar, humanitarian and administrator.
	The theme of the symposium, {\em  Five Decades of Weak
	Interactions}$\cite{5-Decades}$, emphasized his abiding interest
	in chirality and $V-A$.$\cite{V-A}$

}

\section{The Pion at $T = 0$}
	The ubiquitous pion has played a central role in the early history
	of particle physics.  Marshak's two meson theory brilliantly resolved
	the early puzzle and confusion between the ($\mu$) meson
 	observed at sea level and the strongly interacting $\pi$ meson
	proposed by Yukawa.  The two meson theory was later confirmed by
	accelerator experiments that produced copious numbers of
	pions which decay to muons.

	How does the pion couple to matter?

	Theorists struggled with this.  Early on, there was a dichotomy
	in the description, whether it should be a simple $\gn$ coupling
	or a derivative coupling.  One was a renormalizable coupling,
	while the other was manifestly not.  And yet from the point
	of view of pion coupling to nuclear matter, there was an
	equivalence theorem that was valid in the nonrelativistic
	limit.

	With the advent of current algebra and PCAC, the realization grew
	that the pion coupling is a special one.  Rather than a simple
	$\gn$ coupling, it is fundamentally a derivative coupling to matter.

	The reason for this derivative coupling was clarified when
	the work of Nambu-Jona-Lasinio (NJL)$\cite{NJL}$ led to a new
	understanding of the
	special role that the pion would play in the world at $T=0$.
	For the fundamental Lagrangian, in the absence of a primordial mass,
	has the $U(2)_{_{A}}$ chiral symmetry
\be
   	\psi (\x)  \rightarrow    {\rm e}^{i \,\alpha \gamma_{_{5}} } \;
			{\rm e}^{i \,\vec{\tau} \cdot \vec{\alpha}
                        \gamma_{_{5}} } \psi (\x)
\ee
	This symmetry, so to speak, protects the fermion from acquiring
	a mass, just as gauge invariance in its simplistic form protects
	the photon from acquiring a mass.  The Noether charge
	associated with this chiral symmetry has the form
\barray
   	\Qij	&=&  \frac{1}{2} \; \int d^3 x \; \psidag{}_{j} (\x)
				\gn  \psi^{i} (\x) \\
        	&=& - \frac{1}{2} \sumps \, s \; \left( \adag{}_{j}  \aps{}^{i}
                   - \bps{}_{j}   \bdag{}^{i}   + \delta^{i}_{j}
                   \right)                           \label{Qa-expansion}
\earray
	where $\aps$ and $\bps$ are the massless quark and antiquark
	operators, and $s$ is defined to be $\pm 1$ for $R$
	and $L$ helicities respectively.

	$\Qij$ may be decomposed into the Noether charges, $\Q$ for the
	$U(1)_{_{A}}$ symmetry, and $\Qa$ for the $SU(2)_{_{A}}$
	symmetry.  $\Q$ has, however, an instanton anomaly so that it
	is not a constant of motion
	but the isovector charge $\Qa$ commutes with the
	Hamiltonian
\be
   	[  H,  \Qa ]  = 0.
\ee
	Naively, we would expect the ground state to be invariant
	under $\Qa$ as well.
	As Nambu and Jona-Lasinio have already pointed out in 1961, however,
	the internal dynamics could result in a ground state that
	nevertheless is not invariant.  This
	is much like in the ferromagnetic analogy, where even though the
	Hamiltonian is rotational invariant, the ground state
	is not the $J=0$ state.
	It is instead a state where all the $N$ spins are lined up,
	so that it has maximum eigenvalue, $N \hbar/2$.

	As a result of the dynamical symmetry breakdown, the fermions
	acquire mass, even though the fermions in the fundamental Lagrangian
	are massless.
	In addition, as Nambu and
	Goldstone$\cite{Goldstone,Goldstone-Salam-Weinberg}$
	have shown, a signature of this dynamical symmetry
	breakdown is the existence of a strictly massless
	excitation carrying the quantum numbers of $\Qa$.

	But the observed pions are not massless, indicating an explicit
	breaking of the chiral symmetry.  This
	is as a result of the electroweak breaking that generates
	a tree level running mass for the quarks through the Yukawa
	coupling.
	The pion is thus only an approximate Nambu-Goldstone boson, and
	we may use the (broken) QCD Noether charge as the interpolating
	field for the pion,
\be
   	\phi^{a} (\x)  =  \frac{i}{m_{\pi}^2 \, f_{\pi}}   \;\partial^{\mu}
			  J^{5 a}_{\mu} (\x)	     \label{pion-field-1}
\ee
	In the limit $m_{\pi} \rightarrow 0$, the axial current is
	conserved, and the interpolating field goes over into
\be
	\phi^{a} (\x)  =  i \frac{a_{\pi}}{\Lambda_c}  \;
			\bar{\psi} \; \gn
			\tau^{a} \; \psi             \label{pion-field-2}
\ee
	The two forms of the interpolating fields are equivalent,
	although for low energy phenomenology, the former is more
	convenient.
	Because of the derivative in the interpolating field
	eq.(\ref{pion-field-1}), it is easy to see directly that the
	pion interaction with matter obeys the zero energy decoupling
	theorem.  A soft pion with $p_{\mu} \rightarrow 0$
	decouples from matter.

\section{Probing the Vacuum}

	As Lee$\cite{Lee-75}$ empahsized in 1975, the physical
	vacuum instead of being an `empty' state is rich with structure.
	In our case, the NJL vacuum is filled with
	massless quark-antiquark pairs of the same helicity.
	How may we hope to probe the vacuum and measure this rich
	structure?

	One way is to create a mini big-bang at a heavy ion accelerator
	and hope that the violent collision creates in the center of
	mass frame a new vacuum in a small region of space which
	rapidly expands.  Such a new vacuum will be a rich source of
	pion multiplicities which evaporate and may be observed at
	RHIC.

	This new vacuum will not have translational invariance, since
	the center of mass is clearly the preferred frame.
	As suggested by Rubakov {\em  et al}, $\cite{Rubakov-86}$,
	it is conceivable that the system could develop different domains
	within each of which is a different orientation of chiral
	condensates.  These disoriented chiral condensates$\cite{dcc}$
	lead to new charge correlations in the pion multiplicities.

	The other way is to probe the physical vacuum through the
	measurement of order parameters.  For chiral symmetry, the
	well-known order parameter is $\psibarpsi$.
	There is a famous theorem which states that if $\psibarpsi
	\neq 0$, then
\be
   	\Qa   \vac    \neq 0.
\ee

	This is easily seen to be an immediate consequence of the
	the equal time commutation rule
\begin{equation}
     	[ \Qa \; , \;\; i \, \int d^3 x \; \psibar \, \tau^{b} \,\gn \, \psi ]
              	=  - i \, \delta^{ab}  \,\int d^3 x \; \psibar
                \; \psi                             \label{Q5-algebra}
\end{equation}
	since by taking the vacuum expectation of the operators
	on both sides, a non-vanishing $\psibarpsi$ implies
	that $\Qa$ cannot annihilate the vacuum.

	At issue in our later discussion is whether the converse
	of this theorem should also be true.  I will show here by
	explicit construction that the converse is indeed false,
	and point out how $\psibarpsi$ is an incomplete indicator
	of dynamical chiral symmetry breaking in the ground state.

	To fully probe the structure of the vacuum, other order
	parameters will have to be measured.  As we shall see below,
	they are non-local order parameters, which explains why they
	have not been considered before by others.

\section{NJL vacuum with Flavor}

	Nambu and Jona-Lasinio worked with the case of one flavor in writing
	down the paired state.  They worked with a relativistic generlization
	of the BCS theory in superconductivity.
	The NJL ground state may be obtained as an $X_2$ rotation of the
	Fock space vacuum$\cite{Chang-hisig}$
\be
   	\vac    =  \prodps {\rm e}^{i \,\thetap \,X_{2p}}  \, | 0 >
\ee
	where $X_{2p}$ is an element of the chirality algebra, represented
	by
\barray
     	X_{3p}	&=&	- \sum_s \; \frac{s}{2}  \left( \adag \aps
			+ \bdag \bps \right) \\
	X_{2p}	&=&	\,i \,\sum_s \; \; \frac{s}{2}  \left( \adag \bdag
			- \bps \aps \right) \\
	X_{1p}	&=&	\;\; \,\sum_s \; \frac{1}{2} \left( \adag \bdag
			+ \bps \aps \right)
\earray
	The $X$ operators generate the $SU(2)$ algebra, with $X_{3p}$
	easily recognizable as the usual chirality charge.
	The $X_{2p}$ rotation populates the Fock vacuum with quark-antiquark
	pairs.
	The angle, $\thetap$, of this rotation is related to the mass $m$
	acquired by the fermion through the NJL gap equation
\be
   	\tan{2\thetap}  =  \frac{m}{p}   	     \label{gap-eqn}
\ee
	Note that the limiting angle, $\thetap = \pi /4$, corresponds
	to an infinite mass gap.

	For strong interaction dynamics, we should in principle work with QCD.
	The ground state here involves not just quarks and antiquarks but
	also gluons.
	We imagine working with an effective theory where the gluon degrees
	have been integrated out, and so will continue to deal only with
	quarks and antiquarks.

	For the case of two flavors, we shall consider the generalization
	of the NJL vacuum.  We begin with the state
\be
   	\vac{}^{\prime} =  {\rm e}^{i \,\thetap X_{2p}}
			\; {\rm e}^{i \, \thetapp \;
			\vec{X}_{2p} \,\cdot \hat{n}} \;\; | \Omega >
\ee
	where $\vec{X}_{2p}$ is an obvious isovector generalization
	of $X_{2p}$
\be
	\vec{X}_{2p}	=	\,i \; \sum_s\; \frac{s}{2}  \left( \adag
				\,\vec{\tau} \,\bdag
			- \bps \,\vec{\tau} \,\aps \right)
\ee
	and $| \Omega >$ is the $\Qa$ chiral invariant background state.
	This background state is itself populated with {\em quartet }
	of quarks and antiquarks
\be
   	| \Omega >   =  \prodps  \left( \cos{\alpha} \;+\; \sin{\alpha}
			\;a^{\dagger}_{1,p,s} a^{\dagger}_{2,p,s}
			b^{\dagger 1}_{-p,s} b^{\dagger 2}_{-p,s}
			\;\right)  | 0 >
\ee
	In general, for $\alpha$ not equal to the special fixed point,
	$\alpha = \pi /4$, the state $\vac{}^{\prime}$ is dependent
	on $\thetap{}^{\prime}$, and it is not invariant under
	$\Qa$ chirality.  For the asymptotic case of $\alpha = \pi /4$,
	however, the background state is unchanged under the
	$\vec{X}_{2p}$ rotation, and  $\vac{}^{\prime}$ becomes
	$\Qa$ invariant.

	From this, we may construct the most general representation
	of a chiral non-invariant NJL vacuum by writing
\barray
        \vac  	&=&  \prodps \left\{ f -  \;s\; \bar{g} \, {\rm e}^{i\xi}
\,\adag
		- \;s\; \bar{g}^{\prime}\,
		{\rm e}^{i\xi}\,\hat{n} \cdot (\adag
		\vec{\tau} \bdag)  \right. \nonumber \\
		&&    \left.  \;\;\;\;\;\;
		+ h \, {\rm e}^{2i\xi}\, a_{1,p,s}^{\dagger}
                                         a_{2,p,s}^{\dagger}
		    b_{-p,s}^{\dagger 1} b_{-p,s}^{\dagger 2}  \right\}
                 | 0 \rangle                           \label{NJL-vac-T}
\earray
	where
\barray
     	f 	&=&	\frac{\cos{\alpha}}{2} \left( \costthetap
			+ \costthetapp \right)
			+ \frac{\sin{\alpha}}{2} \left( \costthetap
			- \costthetapp \right) \\
	h	&=&	\frac{\cos{\alpha}}{2} \left( \costthetap
			- \costthetapp \right)
			+ \frac{\sin{\alpha}}{2} \left( \costthetap
			+ \costthetapp \right)  \\
	\bar{g}\; &=&	\;\;\;\left( g\cosxip  +
			i g^{\prime} \sinxip \right)   \\
	\bar{g}^{\prime} &=&	\;\;\;\left( g^{\prime} \cosxip +
			i g \sinxip \right)    \\
	g\;	&=&	\frac{(\cos{\alpha} + \sin{\alpha})}{2}
			\sintthetap      \\
	g^{\prime}	&=&	\frac{(\cos{\alpha} - \sin{\alpha})}{2}
			\sintthetapp
\earray
	The generalization here is to introduce the $\xi$ and
	$\xip$ phases.  At $T=0$, we may make use of the
	arbitrary phase in the $\bps$ operator to redefine it so that
	the $f, g, g^{\prime}$ and $h$ functions
	are real.   This comes about because the $\bps$ operator
	is defined in terms of the filled Dirac sea, and there is an
	arbitrary phase in the relation between $\bps$ and the negative
	energy operator, $d^{\dagger}_{p,s}$. Once this has been fixed,
	however, the evolution of the functions for higher $T$
	may lead to new phases that cannot be absorbed into the
	definition of $\bps$.  So, our eq.(\ref{NJL-vac-T}) is
	meant for the general thermal vacuum at higher $T$.

	Under a chiral transformation, the NJL vacuum
	changes to a new unitarily inequivalent ground state.
	Since $\Qa$ commutes with the Hamiltonian, this new ground state must
	have the same energy as the original one. It costs nothing in energy to
	chiral transform the vacuum.  The excitations associated
	with this transformation are thus zero energy (Nambu-Goldstone)
	modes.  For the case of two flavors, the Nambu-Goldstone modes
	have the same quantum numbers as the $\vec{\pi}$ triplet.

\section{Order Parameters Old \& New}
	The standard signal for the breakdown of chiral symmetry
	is the nonvanishing of $\psibarpsi$.  As we shall see in this
	section it is a poor indicator of the state of chiral
	symmetry of the vacuum.  A vanishing $\psibarpsi$ does not
	imply that the vacuum is chiral symmetric.

	To see this, we may evaluate explicitly the standard order
	parameter and find
\barray
   	\frac{1}{2} \psibarpsi 	&=& 	\frac{1}{V}
				\sumps \; \frac{s}{2} \; \langle vac |
				\left( \adagbdag \;+\; \bps \aps \right)
				| vac \rangle  \\
			&=&  	- \frac{\cos{2\alpha}}{V} \sumps
				\left( \;\sintthetap
				\costthetapp
				\cosxi \cosxip
				\;-\; \costthetap
				\sintthetapp
				\sinxi \sinxip \;
				\right)		\label{psibarpsi-T}
\earray
	Eq.(\ref{psibarpsi-T}) shows that, contrary to popular
	folklore, a vanishing $\psibarpsi$ {\em  does not necessarily}
	imply that the vacuum is chirally symmetric.
	For example, if the phases $(\xi,\xip) = (\pi/2, 0)$, say,
	the order parameter would vanish even though the vacuum has a non-zero
	chirality angle, $\thetapp$. Eq.(\ref{NJL-vac-T}) shows that
	the corresponding NJL vacuum is not $\Qa$ invariant.

	$\psibarpsi$ by itself is thus a deficient indicator of the
	state of chiral symmetry at high temperatures.
	We need to measure other order parameters.
	Actually there is an $SU(2N_f) \times SU(2N_f)$
	chirality algebra$\cite{chiral-alg}$ formed out of the bilinear
	quark and antiquark operators.  The algebra includes the order
	parameters identified as
\barray
	Y_{p} 		&=&	\;\;\sum_{s} \;\frac{s}{2} \;
				\left( \adagbdag  +  \bps \aps \right)  \\
	\vec{Y}_{p}	&=&	\;\;\sum_{s} \;\frac{s}{2} \;
				\left( \adag \,\vec{\tau}\, \bdag
				+  \bps \,\vec{\tau}\, \aps \right)  \\
	X_{2p}		&=&	i \, \sum_{s} \;\frac{s}{2} \;
				\left( \adagbdag  -  \bps \aps \right)  \\
	\vec{X}_{2p}	&=&	i \, \sum_{s} \;\frac{s}{2} \;
				\left( \adag \,\vec{\tau}\, \bdag
				-  \bps \,\vec{\tau}\, \aps \right)  \\
	\vec{N}_{p}	&=&	\;\;\sum_{s} \; \frac{1}{2} \;
				\left( \adag \,\vec{\tau}\, \aps
				- \bps \,\vec{\tau}\, \bdag \right)
\earray
	where $\psibarpsi$ may be obtained by a sum over all momenta
	of $2 Y_{p}$.

	By taking the vacuum expectation values of these order parameters
	in the generalized NJL vacuum, it can be shown that
	{\em  only when the vacuum expectation value of all five order
	parameters listed above vanish} may we conclude
	that the ground state is chirally symmetric.$\cite{chiral-alg}$

	While $Y_{p}$ and $\vec{Y}_{p}$, when summed over $\vecp$, are
	related to the familiar local order parameters in space-time, the
	other order parameters are new and intrinsically non-local, with
\barray
	\sum_{p} X_{2p}		&=&	\frac{i}{4} \int d^3 x \; \bar{\psi}
				\frac{ \gamma_o \vec{\gamma} \cdot
                                \vec{\nabla}}
				{\sqrt{- \nabla^2}}  \psi  \;+\;
				{\rm h.c.}\\
	\sum_{p} \vec{N}_{p}	&=&	\frac{1}{4} \int d^3 x \; \bar{\psi}\;
				\vec{\tau} \;\frac{\vec{\gamma} \cdot
                                \vec{\nabla}}
				{\sqrt{- \nabla^2}}  \psi  \;+\; {\rm h.c.}
\earray
	Their vacuum expectation values yield new insights into the
	structure of the chiral broken vacuum.

\section{Signatures at high $T$}

	QCD studies have shown that $\psibarpsi$ vanishes at $T_c$,
	and stays zero above $T_c$.  The popular folklore is to conclude
	that chiral symmetry is restored above $T_c$.
	This is supported by the work of Tomboulis and Yaffe$\cite{Tomboulis}$
	who have shown rigorously in lattice gauge theories that the effective
	action has a strict global chiral invariance at high $T$.
	This is also seen in the manifest chiral symmetry of the continuum
	QCD effective action at high $T$ as derived by Braaten-Pisarski
	and Frenkel-Taylor-Wong.$\cite{BP}$

	And yet, the same continuum studies also show that a massless
	quark propagates in the hot QED and QCD environ as if it had
	a mass gap proportional to
	$gT$, for $p$ large.$\cite{Weldon-Klimov,Donoghue-Chang-hiT-Barton}$
	At zero $T$, the origin of fermion mass was attributed to chiral
	symmetry breakdown.
	Could this thermal mass at high $T$ be due to the continued
	breakdown of chiral symmetry?  If so, how is it possible that the
	effective action for hot QCD is nevertheless chiral invariant?

	To resolve the apparent paradox in these signatures,
	it is best to follow the evolution of the NJL vacuum
	with temperature.  In this picture, the key to
	chiral symmetry breaking is in the population of
	quark-antiquark pairs in the ground state.  This
	population is controlled by the chirality angles,
	$\thetap$.

	In this language, the popular folklore asserts that $\thetap$
	should decrease with temperature, so that the ground state
	becomes depopulated of these pairs.
	The pairs `heat up' with temperature and proceed to break up.
	When the temperature reaches a critical point, $\thetap$ vanishes,
	and there are no more pairs in the new thermal ground state,
	and chiral symmetry is said to be restored.
	In the popular jargon, the quark-antiquark condensate dissolve
	above $T>T_c$, and it very nicely explains away the vanishing of
	$\psibarpsi$.

	According to eq.(\ref{gap-eqn}), when $\thetap = 0$,
	the gap parameter vanishes, and the quark should therefore
	propagate in a hot medium as if it were massless.  For the NJL model,
	this is indeed what happens.

	For QCD, however, the scenario with a vanishing mass gap
	poses a problem.  For it conflicts with the well-known
	result that the massless quark propagates in a hot QCD
	medium as if it had a `thermal mass' proportional  to $T$.

	Could it be that QCD opts for an alternate scenario?
	As temperature increases, the chirality angle $\thetap$
	{\em  increases} rather than decreases, so that it
	approaches the critical angle $\pi /4$ as $T \rightarrow \infty$.
	In this limit, according to eq.(\ref{gap-eqn}) the gap parameter
	approaches infinity proportional to $T$, thus explaining the
	`thermal mass'.

	At first sight, you might worry that a vanishing $\psibarpsi$
	would pose a problem for this scenario.
	But as eq.(\ref{psibarpsi-T}) shows, a generalized NJL vacuum
	may very easily be populated with quark-antiquark pairs and
	still enjoy a vanishing $\psibarpsi$.

	Such a thermal vacuum continues to
	break the old zero temperature chirality, and the pion should
	remain strictly massless at high $T$.
	This conclusion conforms to the earlier continuum study of the
	effective QCD action with external scalar $j$ and pseudoscalar
	$j_{_{5}}$ sources which showed that the QCD pion is massless for
	all $T$.$\cite{Chang-QCD-pion}$

	But in spite of the continued breaking of zero temperature chirality,
	the QCD effective Lagrangian at high $T$$\cite{BP}$
\be
   {\cal L}_{\rm eff} = - \bar{\psi}_{\beta} \gamma_{\mu} \partial^{\mu}
                          \psi_{\beta}
                       - \frac{\Tprime {}^2}{2} \, \bar{\psi}_{\beta}
                         \left<
                       \frac{\gamma_o - \vec{\gamma} \cdot \hat{n} }
                       {D_o + \hat{n} \cdot \vec{D} }
                         \right> \psi_{\beta}          \label{BP-pspace}
\ee
	shows a manifest global chiral symmetry invariance
\be
	\psibeta    \longrightarrow    {\rm e}^{i \beta \gn } \; \psibeta
\ee
	How may we reconcile the two?

\section{Chiral metamorphosis}

	The answer comes when
	you examine the Noether charge for this high $T$
	chiral symmetry.  It is a {\em  metamorphosis} from the
	$T=0$ Noether charge, $\Qa$.  By carrying out a
	canonical quantization of this non-local effective
	action$\cite{xcurrent}$, I have found that it is given by
\be
   	\Qabeta  =  - \frac{1}{2} \sumps \, s \; \left( \Apsdag
			\,\tau^{a} \,\Aps
                   - \Bps  \,\tau^{a} \, \Bpsdag
                   \right) 			     \label{Qa-beta}
\ee
	to be compared with the original chiral charge
\be
	\Qa      =  - \frac{1}{2} \sumps \, s \; \left( \adag
			\,\tau^{a} \,\aps
                   - \bps  \,\tau^{a} \, \bdag
                   \right)			     \label{Qa-expansion-1}
\ee
	The curious thing about this high temperature chirality is
	that the operators $\Aps$ and $\Bps$ in eq.(\ref{Qa-beta}) here
	are {\em  massive} operators.  They are
	in contrast to eq.(\ref{Qa-expansion-1}), given in terms of {\em
	massless} operators.

	The new thermal vacuum is invariant under the high $T$ chirality.
	It is annihilated by $\Qabeta$  through the defining
	properties
\be
   	\Aps   \vacbeta    \;=\; \Bps   \vacbeta
			\;=\;  0     	             \label{vac-T-prop}
\ee
	As we will show below this thermal vacuum
	however is not annihilated by the
	massless operators, $\aps$ and $\bps$.
	And by eq.(\ref{Qa-expansion}), we may conclude that
	the zero temperature $\Qa$ symmetry is not restored for $T > T_c$.
	Instead, what happens is a metamorphosis of the chiral
	symmetry across $T_c$.

	{\em  The morphing consists in replacing the massless with the
	massive operators in eq.(\ref{Qa-expansion}).}

\section{Free vs Dressed Vacuum}

	How is this new thermal vacuum related to the vacuum of the
	massless free field?

	At this point, a word about the notion of a thermal vacuum may
	be in order.  The thermal average is taken over the
	entire spectrum of states with the usual Boltzmann weight.
	In thermofield dynamics$\cite{Umezawa}$, or real time field
	theory$\cite{CTP}$, we replace this thermal average with a
	thermal vacuum expectation value
\barray
        <  \bar{\psi}_{_{H}} (0)  \psi_{_{H}} (0)  >_{av}
        	&=& \frac{1}{Z}\;  \sum_{n} {\rm e}^{-\beta E_n} \;
                        < n |  \bar{\psi}_{_{H}}(0)  \psi_{_{H}}
			(0) | n >  \\
		&\equiv& {}_{_{\beta}} \langle vac |
			 \psibarbeta (0) \psibeta (0)
			| vac \rangle_{_{\beta}}
\earray
	where $\vacbeta$ is a doubled Hilbert space, supplemented with
	the operators of the heat bath.  The $\tilde{}$ heat bath degrees
	of freedom are a mirror image of the physical degrees of freedom, with
	the crucial difference that the time evolution operator is
	governed by $- \tilde{H}$.

	Recall that in the original Hilbert space in which $\psi_{_{H}}$
	operates, at time $t=0$, the Heisenberg field overlaps with the
	asymptotic field, $\psi_{in}$, as well as with the perturbative free
	field, $\psi_{o}$.  From the local field identity
\be
   	\bar{\psi}_{o} \psi_{o} (0)   =
			\bar{\psi}_{_{H}}  \psi_{_{H}} (0)
\ee
	it follows that
\be
   	< \bar{\psi}_{o} \psi_{o} >_{av}    =
				< \bar{\psi}_{_{H}} \psi_{_{H}} >_{av}
\ee
	and since at high $T$ the order parameter vanishes, we arrive at
	the condition in terms of the free massless operators
\be
   	\frac{1}{V} \sumps \;\,{}_{_{\beta}} \langle  vac |
		\;s\, \left( \adagbdag +  \bps \aps \right)  \; \vacbeta
			= 0			     \label{thermal-bc}
\ee
	This equation provides the boundary condition that relates the
	thermal vacuum $\vacbeta$, defined by eq.(\ref{vac-T-prop}),
	to the perturbative free vacuum, defined
	by
\be
	\aps  | 0 >   =  \bps  | 0 >
\ee
	This boundary condition is in fact what we have written down in
	eq.(\ref{psibarpsi-T}), with the added proviso that the mass
	gaps are proportional to $T$, since $\Aps$ and $\Bps$ are massive
	operators.

\section{Is there more than one $T_c$?}

	The BP-FTW effective action for hot QCD is valid at high $T$
	when chirality has already morphed into the new $\Qbeta$ phase.
	Without studying the transition taking place at $T_c$,
	it is not possible to say definitively the precise phase that
	the  vacuum is in.  It is nevertheless interesting to speculate
	on it.

	At zero temperature, the phase angles $\xi$ and $\xip$ are
	vanishing, as per definition, while the chirality angles,
	$\thetap$ and $\thetapp$, corresponding to the isoscalar
	and isovector mass gaps, are non-zero.
	In the presence of the heat bath, the angles will become
	temperature dependent.  For the chirality angles, they
	may range between $0$ and $\pi /4$, corresponding to
	the physical range of the mass gap.  At any finite
	temperature, the chirality angles must be {\em  less than}
	$\pi /4$.

 	The new $\Qbeta$ phase will set in at the
	transition point due to any of the following
	possibilities: \\
	\begin{enumerate}
     \item
		For $T \geq T_{c}$, $\;\;(\xi, \xip) = (
		\frac{\displaystyle \pi}{\displaystyle 2}, 0)$
		or $\;\;(0,  \frac{\displaystyle\pi}{\displaystyle 2} )$  \\

		I like this possibility the most because of the mysterious
		phase of $\pi/2$ that it involves.
		The $\xi$ and $\xip$ start at zero as temperature rises,
		until one of them reaches $\pi /2$ and freezes there for
		$T \geq T_c$. The chirality angles, $\thetap$ and
		$\thetapp$ begin at some non-vanishing values
		corresponding to the chiral broken ground state
		isoscalar and isovector mass gaps.  These mass gaps
		grow with temperature and continue through the $T_c$
		to become proportional to $T$ at high $T$.
		This scenario applies to any number of generations.  \\

     \item
		For $T \geq T_c$, $\;\;(\xi, \xip) = ( 0, 0)\;\;$, with
		$\;\;\thetap = 0$  \\

		In this scenario, the phase angles $\xi$ and $\xip$
		remain zero throughout, while $\thetap$ decreases with
		temperature
		until it reaches zero at $T_c$ and stays zero for
		higher temperatures.  $\thetapp$, on the other hand, grows
		with temperature.
		Above $T_c$, the $\Qa$ chiral
		symmetry is broken by the isovector mass gap which
		becomes proportional to $T$ at high $T$.
		This scenario is possible only for two or more generations.\\
		Actually, eq.(\ref{psibarpsi-T}) suggests also the alternative
		possibility of having $\thetapp = \pi/4$ at $T_c$
		in lieu of $\thetap = 0$.
		This is ruled out, however, on physical grounds.  For
		if $\thetapp$ reaches $\pi /4$ at $T_c$, the
		mass gap will blow up at a finite temperature.  \\

     \item
                For $T \geq T_c$, $\;\;(\xi, \xip) = (
		\frac{ \displaystyle \pi}{ \displaystyle 2} ,
		\frac{ \displaystyle \pi}{ \displaystyle 2} )\;\;$,
		with $\;\;\thetap = 0$ \\

		In this scenario, both $\xi$ and $\xip$ reach $\pi /2$
		at the same temperature, $T_c$, when the isoscalar
		mass gap vanishes.  The isovector mass gap increases
		with temperature as before.  \\
	\end{enumerate}

	With such a rich structure of phases, it would be surprising that
	they all undergo the phase transition at the same temperature,
	$T_c$.  This raises the interesting question as to whether there
	are more than one $T_c$ in the so-called transition region,
	with $T_{c1}$ as the temperature where $\xi$ reaches $\pi/2$,
	and $T_{c2}$ the temperature when $\xip$ reaches $\pi/2$.
	Studies of the new order parameters promise to shed more
	light on the chiral morphing transition region.

\section{The Pion in the Early Universe}

	So far, our attention has been focussed on the structure of the
	QCD vacuum at high temperatures.  But the real messenger of
	the broken vacuum is the pion.  At zero temperature, the
	properties of the pion have been well understood.  At high
	temperatures, how does the pion interact with matter?

	As mentioned earlier, the continued breaking of the old
	$\Qa$ chirality demands that the pion be a Nambu-Goldstone boson
	at high $T$.  In the standard model, with a fundamental Higgs,
	electroweak symmetry is restored above $T_{ew}$.
	There is thus no explicit quark mass term in the effective Lagrangian
	at high $T$.  The continued breaking of $\Qa$ chiral symmetry
	implies a {\em  strictly massless} pion at high $T$.

	The implications for this for the early universe are
	profound.

	In the standard folklore, the early universe is an alphabet soup
	of quarks and gluons.  At first people thought that the quarks
	and gluons are all massless.  This has since been qualified with
	the realization that the quarks and gluons had thermal mass.
	It being of order $gT$, the thermal mass does not have a major
	impact on the equipartition of energy between quarks and gluons.

	What is significant in the standard folklore is that the pions
	are missing from the alphabet soup.

	In our understanding of the state of chiral symmetry at high $T$,
	the pion will have a very good reason to be in the alphabet
	soup in the early universe.  Namely, the Nambu-Goldstone theorem
	is going to {\em  demand} that the binding energy of the
	massless quark-antiquark state be so deep that the resulting
	bound state must be strictly massless.  They cannot dissociate
	with temperature.

	To be sure, the presence of pions in the alphabet soup is not
	going to change critically the entropy of the early universe,
	since they do not have that many degrees of freedom.  There
	will presumably be other more subtle changes in the scenario
	of the cooling of the early universe.  Not being a cosmology
	expert, I can only speculate that the ubiquitous pion will
	continue to hold the secret to the behavior of matter at
	high temperature and high densities.

	I will end this discussion with a note on the properties of the
	pion at high temperature.  For while the pion propagator has a
	massless pole in the complex momentum plane, lattice calculations
	also show that the pion has a {\em  screening mass} that is
	proportional to $T$.

	Is there a conflict between the two results?

	What I will show is that in a hot environ, you can have a particle
	propagating through the medium possessing both those properties.
        In the language of real time thermal field theory, it is easy
        to find an example of such a particle.  For the
        physical massless pole is determined from the condition that
        the denominator of the propagator vanish$\cite{pion-halo}$
\be
        \Gamma^{(2)}_{\pi} (p, p_o, T) = p^2 ( 1 + {\cal A} )^2 - p_o^2
                                     ( 1 + {\cal B} )^2 = 0
\ee
        where ${\cal A}$ and ${\cal B}$ are functions of $p, p_o, T$.
        The screening mass on the other hand comes from integrating over
        the $x, y, t$ coordinates (i.e. set $p_x = p_y = p_o = 0$)
        in the propagator, so that the pole for the correlation
        function in $z$ occurs at $p_z = i m_{sc}$, where
        $1 + {\cal A} ( im_{sc}, 0, T)  = 0$
        In terms of a physical picture, when we receive light from a
        charged particle, we see it at its retarded position, and
        it is a sharp image.  For the pion, the retarded function
        reads
\be
        D_{\rm ret} (\x) = \theta (-t) \left\{ \delta(t^2 - r^2)
                    + \frac{T}{r} \theta(t^2 - r^2)
                    \left[ {\rm e}^{-T | t-r| }
                    +  {\rm e}^{-T | t+r| } \right] \right\}
\ee
        so that the screening mass leads to an accompanying modulator
        signal that `hugs' the light cone, with a screening length
        $\propto 1/T$.

\section{Acknowledgment}
	Much of the work reported in this article has been done in collaboration
	with Lay Nam.  The work really started as part of a US-China cooperative
	project$\cite{Chang-hisig,Chang-HK}$ with K.C. Chou, the expert on Closed Time
	real time thermal field theory.  The work has spanned many years, and
	it is a pleasure to acknowledge the hospitality of Professor Zimmerman
	at Max Planck Institute in Munich '92, and the hospitality in '93 of
	Professor S.C. Lee at the Institute of Physics in Academia Sinica,
	Republic of China, where various parts of this work was done.

\end{document}